\documentclass[aps,prd,reprint,longbibliography]{revtex4-1}

\usepackage[english]{babel}
\usepackage[utf8]{inputenc}
\usepackage[hyperindex,breaklinks]{hyperref}
\usepackage{amsmath}
\usepackage{amssymb}
\usepackage{float}
\usepackage{graphicx}
\usepackage{placeins}
\usepackage{caption}
\usepackage{subcaption}
\usepackage{bm}
\DeclareMathOperator*{\argmin}{arg\,min}

\renewcommand{\cite}{\citep}

\begin{document}
\title{Optimisation based algorithm for finding the action of cosmological phase transitions}
\date{\today}
\author{Michael Bardsley}
\affiliation{School of Physics and Astronomy, Monash University, Melbourne, VIC 3800, Australia}
\affiliation{Institute of Physics, Laboratory for Particle Physics and Cosmology,
École Polytechnique Fédérale de Lausanne, CH-1015 Lausanne, Switzerland}

\begin{abstract}
We present the OptiBounce algorithm, a new and fast method for finding the bounce action for cosmological phase transitions. This is done by direct solution of the ``reduced" minimisation problem proposed by Coleman, Glaser, and Martin. By using a new formula for the action, our method avoids the rescaling step used in other algorithms based on this formulation. The bounce path is represented using a pseudo-spectral Gauss-Legendre collocation scheme leading to a non-linear optimisation problem over the collocation coefficients. Efficient solution of this problem is enabled by recent advances in automatic differentiation, sparse matrix representation and large scale non-linear programming. The algorithm is optimised for finding nucleation temperatures by sharing model initialisation work between instances of the calculation when operating at different temperatures. We present numerical results on a range of potentials with up to 20 scalar fields, demonstrating $O(1\%)$ agreement with existing codes and highly favourable performance characteristics.
\end{abstract}

\maketitle

\section{Introduction}

% Physics context 
Scenarios in which one or more scalar fields undergo a first order phase transition in the early universe possess a rich phenomenology. Models with extended Higgs sectors can exhibit first order phase transitions at the electroweak scale \cite{kakizakiGravitationalWavesProbe2015, balazsGravitationalWavesALIGO2017,papaefstathiouElectroWeakPhaseTransition2020, alvesColliderGravitationalWave2019, athronStrongFirstorderPhase2019a, jiangImpactComplexSinglet2016, dorschSecondHiggsDoublet2017}, leading to observable consequences such as the generation of baryon asymmetry \cite{kuzminAnomalousElectroweakBaryon1985, shaposhnikovBaryonAsymmetryUniverse1987, morrisseyElectroweakBaryogenesis2012, whitePedagogicalIntroductionElectroweak2016} and detectable gravitational waves \cite{wittenCosmicSeparationPhases1984, hoganGravitationalRadiationCosmological1986, cuttingGravitationalWavesVacuum2018, capriniDetectingGravitationalWaves2020}. At the QCD scale first order phase transitions may be implicated in the creation of intergalactic magnetic fields \cite{siglPrimordialMagneticFields1997, tevzadzeMagneticFieldsQCD2012, ellisIntergalacticMagneticFields2019b}. There has also been speculation regarding a landscape of metastable vacua in string-motivated scenarios, creating interest the study of transitions where the number of involved scalar fields is very large \cite{aazamiCosmologyRandomMultifield2006, greeneTumblingLandscapeEvidence2013, dineTunnelingTheoriesMany2015a, masoumiVacuumStatisticsStability2016a}. 

%In models with extended Higgs sectors \cite{kakizakiGravitationalWavesProbe2015, balazsGravitationalWavesALIGO2017,papaefstathiouElectroWeakPhaseTransition2020, alvesColliderGravitationalWave2019, athronStrongFirstorderPhase2019a, jiangImpactComplexSinglet2016, dorschSecondHiggsDoublet2017} first order phase transitions may lead to observable consequences such as the generation of baryon asymmetry \cite{kuzminAnomalousElectroweakBaryon1985, shaposhnikovBaryonAsymmetryUniverse1987, morrisseyElectroweakBaryogenesis2012, whitePedagogicalIntroductionElectroweak2016} and detectable gravitational waves \cite{wittenCosmicSeparationPhases1984, hoganGravitationalRadiationCosmological1986, cuttingGravitationalWavesVacuum2018, capriniDetectingGravitationalWaves2020}.

%The bounce equation
Estimating the decay rate $\Gamma$ of the false vacuum is a core computation in the study of cosmological phase transitions. At the nucleation temperature $T_N$ the decay probability per Hubble volume approaches unity and expanding bubbles of the new phase appear. The relevant physical consequences - i.e baryon asymmetry, gravitational wave production and magnetic field creation - depend on $T_N$, so it is necessary to fix this quantity by computing $\Gamma$ across a range of temperatures. In the semiclassical approach due to Coleman \cite{colemanFateFalseVacuum1977}
\begin{equation}
\Gamma = A e^{-S_E}(1 + O(\hbar)),
\end{equation}
where $S_E$ is the Euclidean action 
\begin{equation}
S_E[\bm{\phi}^B(\rho)] = \int_0^{\infty} d\rho \rho^{D-1}\bigg[\frac{1}{2}|\dot{\bm{\phi}}^B|^2 + V(\bm{\phi}^B, T)\bigg]
\end{equation}
for the $O(4)$ symmetric bubble profile $\bm{\phi}^B$, which is the least-action instanton interpolating the true and false vacua, respectively $\bm{\phi}_T$, $\bm{\phi}_F$. In the above $\rho$ is the radial coordinate from the center of the bubble, $\dot{\bm{\phi}} \equiv \partial \bm{\phi} / \partial \rho$, $V(\bm{\phi}, T)$ is the effective scalar potential, $D$ is the number of spacetime dimensions, and the prefactor $A$ carries a sub-exponential temperature dependence \cite{callanFateFalseVacuum1977, lindeFateFalseVacuum1981}. $\bm{\phi}^B$ is known as the critical bubble or ``bounce", and satisfies the equations of motion:
\begin{equation}
\label{eq:bounce_equation}
\ddot{\phi_i}^B + \frac{D - 1}{\rho}\dot{\phi_i}^B = \frac{\partial V}{\partial \phi_i^B},\ 1 \leq i \leq n_{\phi},
\end{equation}
where $n_{\phi}$ is the number of scalar fields and the boundary conditions are $\dot{\phi}_i^B(0) = 0$, $\lim_{\rho\to\infty}\phi_i^B(\rho) = 0$. Defining $T_C > T_N$ as the critical temperature at which the true and false vacuum are degenerate, the nucleation condition becomes \cite{ellisMaximalStrengthFirstOrder2019}:
\begin{equation}
\label{eq:nucleation_condition}
N(T_N) = \int_{T_C}^{T_N} \frac{dT}{T} \frac{\Gamma(T)}{H(T)^4} = 1.
\end{equation} 
This is typically replaced with the approximate condition  \cite[Ch 4.4]{whitePedagogicalIntroductionElectroweak2016}:
\begin{equation}
\label{eq:nucleation_approx}
\frac{S_E}{T} = 170 - 4 \ln{T} \ln{g_*}
\end{equation}
where $g_*$ is the effective degrees of freedom at the relevant temperature scale. The problem of finding the nucleation temperature is thereby reduced to computing $S_E$ across a range of temperatures to find the largest root of equation \ref{eq:nucleation_approx}.

%Existing approaches
In general the above problem must be solved numerically. For single-field scenarios the shooting approach suggested by Coleman \cite{colemanFateFalseVacuum1977} is effective. Higher numbers of fields make the task considerably more challenging. The shooting approach fails due to the increased number of field space directions at the origin $\rho = 0$, and relaxation based algorithms are impeded by the fact that the bounce is always a saddle point \cite{maziashviliUniquenessNegativeMode2003}, rather than a minimum of the action $S_E$. Additional difficulties stem from the thin-wall regime characterised by:
\begin{equation}
\frac{V(\bm{\phi}_F) - V(\bm{\phi}_T)}{|V(\bm{\phi}_F) + V(\bm{\phi}_T)|} \ll 1
\end{equation}
in which the solutions converge towards step functions at $\rho = \infty$.  A testament to the difficulty of this problem is the large number of algorithms \cite{kusenkoImprovedActionMethod1995c,johnBubbleWallProfiles1999,konstandinNumericalApproachMulti2006a,parkConstrainedPotentialMethod2011,
espinosaFreshLookCalculation2019,guadaMultifieldPolygonalBounces2019,piscopoSolvingDifferentialEquations2019,satoSimpleGradientFlow2020,chigusaBounceConfigurationGradient2020} proposed since the bounce method was invented. A subset of these can be found in public codes. \texttt{CosmoTransitions} \cite{wainwrightCosmoTransitionsComputingCosmological2012} uses a path deformation method. \texttt{AnyBounce} \cite{masoumiEfficientNumericalSolution2017} implements a multiple shooting algorithm. \texttt{BubbleProfiler} \cite{athronBubbleprofilerFindingField2019} uses the perturbative algorithm proposed in  \cite{akulaSemianalyticTechniquesCalculating2016}. \texttt{FindBounce} \cite{guadaFindBouncePackageMultifield2020a} implements the polygonal multifield method derived	 in \cite{guadaMultifieldPolygonalBounces2019}, and \texttt{SimpleBounce} \cite{satoSimpleBounceSimplePackage2021} uses a gradient flow technique \cite{satoSimpleGradientFlow2020}. 

%What distinguishes OptiBounce 
The OptiBounce algorithm described in this paper makes use of an approach not yet seen in a public code. The key idea is to begin with a functional that takes a true minimum at the bounce, then find the solution by direct optimisation over the parameters of a discrete representation of the bounce path $\bm{\phi}(\rho)$. Solving the resulting high dimensional non linear optimisation problem with traditional techniques is computationally expensive, which may explain why this approach has not yet seen general use. However recent advances in algorithmic differentiation \cite{anderssonCasADiSoftwareFramework2019}, combined with gradient based optimisation software such as \texttt{IPOPT} \cite{wachterImplementationInteriorpointFilter2006} can lead to significant improvements in the efficiency of large scale nonlinear optimisation. The numerical results we present in this paper demonstrate that this makes the functional optimisation approach to finding the bounce not only practical, but in many cases an order of magnitude faster than the currently available public codes. We expect that this performance boost will enable phenomenological scans of models with high numbers of fields at a level of detail not possible with existing techniques. 

% Brief summary of rest of paper
The plan of the paper is as follows. Section \ref{algorithm} describes the optimisation problem solved in the OptiBounce algorithm and its relation to the bounce solution. A more detailed derivation of some key results is also given in appendix \ref{derivation}. To implement the algorithm, a discrete scheme to represent the bounce solution must be chosen. Section \ref{discretisation} presents a Legendre-Gauss collocation scheme appropriate to the problem. In section \ref{implementation} we give a brief outline of how the \texttt{CasADi} and \texttt{IPOPT} codes were used to realise our numerical results, and present these results in section \ref{results} before concluding in section \ref{conclusion}.

\section{OptiBounce algorithm} 	
\subsection{Direct optimisation approach to finding the bounce solution}
\label{algorithm}
In this paper the bounce solution $\bm{\phi}^B$ is assumed to belong to the class of radially symmetric field profiles $\bm{\phi}: [0,\infty) \rightarrow \mathbb{R}^{n_{\phi}}$ satisfying $\dot{\phi}_i(0) = 0$, $\lim_{\rho\to\infty}\phi_i(\rho) = 0$, $1 \leq i \leq n_{\phi}$, where $n_{\phi}$ is the number of scalar fields and the false vacuum is at $\bm{\phi}_F = 0$. We denote the set of all functions meeting these conditions by $\bm{\Phi}$. Within this set, $\bm{\phi}^B$ is singled out as a stationary point of the Euclidean action:
\begin{align}
S_E[\bm{\phi}(\rho)] &= \frac{1}{2}\int_0^{\infty} d\rho \rho^{D-1}|\dot{\bm{\phi}}|^2 + \frac{1}{2}\int_0^{\infty} d\rho \rho^{D-1}V(\bm{\phi}) \\
&\equiv T[\bm{\phi}(\rho)] + V[\bm{\phi}(\rho)],
\end{align}
where we have partitioned the action into the kinetic and potential functionals $T[\cdot]$ and $V[\cdot]$.  

More specifically, the bounce is defined as the stationary point of \textit{lowest} action. This is always a saddle point \cite{maziashviliUniquenessNegativeMode2003}, which means that directly minimising $S_E$ will not work. Instead, we follow the ``reduced problem" introduced by Coleman, Glaser, and Martin \cite{colemanActionMinimaSolutions1978}. A proof that this recovers the bounce solution is given in appendix \ref{derivation}. For some fixed $V_0 < 0$, we define the level set:
\begin{equation}
\mathbf{\Phi}_{V_0} \equiv \{\bm{\phi} \in \mathbf{\Phi} : V[\bm{\phi}] = V_0\}.
\end{equation}
While the value of $V_0$ is arbitrary, we fix it to $V_0 = -1$ in this work. The first step in the algorithm is to find an element of $\mathbf{\Phi}_{V_0}$. This is done by solving a constrained optimisation problem over the parameters $r_0, \sigma$ :
\begin{align}
\text{minimise}\ &T[\bm{\phi}^A(\rho ; r_0,\sigma)] \\
\text{subject to}\ &V[\bm{\phi}^A(\rho ; r_0,\sigma)] = V_0,
\end{align}
where $\bm{\phi}^A$ is the kink ansatz:
\begin{equation}
\bm{\phi}^A(\rho; r_0,\sigma) = \frac{1}{2}\bm{\phi}_T\big(1 + \text{Tanh}\bigg[\frac{\rho - r_0}{\sigma}\bigg] + \frac{e^{-\rho}}{\sigma}\text{Sech}\bigg[\frac{r_0}{\sigma}\bigg]^2\big).
\end{equation} 
The minimiser $\tilde{\bm{\phi}}^A$ is then used as an ansatz in the trajectory optimisation problem:
\begin{align}
\label{eq:full_problem}
\text{minimise}\ &T[\bm{\phi}(\rho)] \\
\text{subject to}\ &\phi \in \mathbf{\Phi}_{V_0}.
\end{align}
The latter problem is infinite dimensional in the sense that we optimise over all curves in the level set $\bm{\Phi}_{V_0}$. In practice we optimise over the coefficients of the discrete representation described in section \ref{discretisation}, yielding a large but finite dimensional search space. 

The bounce action can then be computed directly. Denoting the optimal value of $T[\cdot]$ by $T_0$, we find (see appendix \ref{derivation}):
\begin{equation}
\label{eq:action}
S[\bm{\phi}(\rho)] = \bigg[\bigg(\frac{2 - D}{D}\bigg) \frac{T_0}{V_0}\bigg]^{\frac{D}{2} - 1} \frac{2 T_0}{D}.
\end{equation}
The field profile corresponding to the bounce can also be obtained via $\bm{\phi}^B(\rho) = \tilde{\bm{\phi}}(\rho/\sqrt{\lambda_*})$, where $\tilde{\bm{\phi}}$ is the minimiser from the optimisation problem \ref{eq:full_problem} and $\lambda_*$ is the optimal value of the lagrange multiplier corresponding to the constraint $V[\bm{\phi}(\rho)] = V_0$. Numerical estimation of $\lambda_*$ is not necessary due to the analytic result:
\begin{equation}
\label{eq:lambda}
\lambda_* = \bigg(\frac{2 - D}{D}\bigg) \frac{T_0}{V_0}.
\end{equation}

\subsection{Discretisation scheme}
\label{discretisation}
To numerically solve the optimisation problem \ref{eq:full_problem}, we use a finite dimensional set of basis functions to represent the candidate solutions in $\mathbf{\Phi}_{V_0}$. This reduces the variational constraint $V[\phi] = V_0$ to a set of algebraic conditions on the basis coefficients. Likewise, the objective function $T[\phi]$ becomes a polynomial in the same coefficients. As suggested in \cite{gargDirectTrajectoryOptimization2011}, before defining the basis we make a change of variables that maps our problem from $\rho \in [0, \infty)$ to $t \in [-1, 1)$ via
\begin{equation}
\rho = \gamma(t) = B \log{\frac{2}{1 - t}}.
\end{equation}
Since we space our grid points evenly in $[-1, 1)$, the factor $B$ controls the clustering of points near the origin in $\rho$-space. For the purposes of our prototype we set $B = 15$ throughout.

We then employ the local Legendre-Gauss collection scheme described in \cite{huntingtonComparisonGlobalLocal2007}. The domain is divided into $N$ finite elements $[t_k, t_{k+1})$, $k = 0,...,N - 1$ of length $h = 2/N$. Within each element, we use internal coordinates $\tau \in (0,1)$ and choose collocation points $\tau_0 = 0$, $\tau_j = P^j_d$, $j = 1,...,d$ where $P^j_d$ is the $j^{\rm th}$ root of the Legendre polynomial of degree $d$. Throughout this work we use $d = 3$ unless otherwise stated. These internal coordinates are related to the external ones by $t_{k,j} = t_k + h\tau_j$. Finally, we include the point $t_{N,0}$ to represent the asymptotic endpoint $\rho \rightarrow \infty$. 

An advantage of this scheme is accurate numerical integration due to the choice of Gaussian quadrature points. With the weights
\begin{equation}
w_i = \frac{1 - \tau_i}{N^2 P_N(\tau_i)^2},
\end{equation}
we have the quadrature rule
\begin{equation}
\int_{0}^{1} f(\tau)d\tau \approx \sum_{i=1}^{N} w_i f(\tau_i),
\end{equation}
which is exact for polynomials of degree $\leq 2d - 1$ \footnote{On the same number of points, the Newton-Cotes quadrature is of exactness $d - 1$.}. Integration over the whole domain is done by summing the quadrature on each element. To represent the field profile $\bm{\phi} : [-1,1) \rightarrow \mathbb{R}^{n_\phi}$, we use an orthogonal basis of Lagrange polynomials:
\begin{equation}
\label{eq:interpolation}
\bm{\phi}_k(\tau) = \sum_{r=0}^d l^d_r (\tau) \bm{\phi}_{k,r},
\end{equation}
where
\begin{equation}
l^d_i(\tau) = \prod^d_{j=0,j \neq i} \frac{\tau - \tau_i}{\tau_i - \tau_j}.
\end{equation}
The $l^d_i$ satisfy $l^d_i(\tau_j) = \delta_{ij}$, so the basis coefficients $\bm{\phi}_{k,j}$ are just the field values at the collocation points. This means that interpolations are easily constructed by sampling at the points $t_{k,j}$.

Continuity of the field profile at the element boundaries is ensured by the imposition of additional algebraic constraints. Equation \ref{eq:interpolation} provides an estimate of the field values at the end of each finite element:
\begin{equation}
\bm{\phi}_k (1) = \sum_{r=0}^d l^d_r(1) \bm{\phi}_{k,r} \equiv \sum_{r=0}^d D_R \bm{\phi}_{k,r}.
\end{equation}
Setting this estimate equal to the field value at the beginning of the next element leads to the continuity constraint:
\begin{equation}
\label{eq:continuity}
\bm{\phi}_{k+1,0} - \sum_{r=0}^d D_r \bm{\phi}_{k,r} = 0,\ k = 0,...,N-1.
\end{equation}
We also discretise the field profile derivatives at the element boundaries, represented by control variables $\mathbf{u}_k$, $k = 0,...,N$. 
At the collocation points, we use a linear interpolation:
\begin{equation}
\mathbf{u}_{k,j} = (1 - \tau_j)\mathbf{u}_k + \tau_k \mathbf{u}_{k+1},\ 0 \leq k < N.
\end{equation}
We can also differentiate equation \ref{eq:interpolation} to obtain estimates of the derivatives:
\begin{equation}
\dot{\bm{\phi}}_{k,j} \approx \frac{1}{h_k} \sum_{r=0}^d \dot{l}^d_r (\tau_j) \bm{\phi}_{k,r} \equiv \frac{1}{h_k} \sum_{r=0}^d C_{r,j} \bm{\phi}_{k,r}.
\end{equation}
Continuity of the field derivatives is then enforced by the collocation constraint:
\begin{equation}
\label{eq:collocation}
h_k \mathbf{u}_{k,j} - \sum_{r=0}^d C_{r,j} \bm{\phi}_{k,r} = 0,\ k = 0,...,N-1,\ j = 1,...,d.
\end{equation}

Having specified the discretisation scheme, it remains to implement the optimisation problem \ref{eq:full_problem}. The discrete form of the condition $V[\phi] = V_0$ is:
\begin{equation}
\label{eq:v0_discrete}
\sum_{k = 0}^{N - 1} h_k \sum_{j = 1}^d w_j \gamma(t_{k,j})^{D-1} \dot{\gamma}(t_{k,j})V(\bm{\phi}_{k,j}) 	- V_0 = 0.
\end{equation}
The complete set of constraints is then equations \ref{eq:continuity}, \ref{eq:collocation} and \ref{eq:v0_discrete} along with the boundary conditions $\bm{\phi}_{N,0} = \bm{\phi}_T$, $\textbf{u}_0 = 0$. We vary the decision variables $\bm{\phi}_{k,j}$ and $\mathbf{u}_k$, $k = 0,...,N - 1$, $j = 0,...,d$ while seeking to minimise the objective functional:
\begin{equation}
\label{eq:objective_discrete}
T[\bm{\phi}] = \frac{1}{2} \sum_{k=0}^{N-1} h_k \sum_{j = 1}^d w_j \gamma(t_{k,j})^{D-1} \dot{\gamma}(t_{k,j}) ||\mathbf{u}_{k,j}||^2.
\end{equation} 

As a final point we note that in the extreme of the thin wall limit, the field profiles approach a step function. Since we use a fixed grid size $h$, there is a point beyond which this discontinuity falls entirely within a single finite element. Step functions are not well approximated by low-degree polynomials, so this can lead to increased convergence time, oscillations and ultimately failure to converge. Increasing the degree of the internal polynomials can mitigate this to an extent. We believe that a better solution would be to implement an adaptive mesh refinement collocation scheme \cite{liuAdaptiveMeshRefinement2015a} which automatically increases the grid resolution in the presence of large derivatives. We intend to address this in a future work. 

\subsection{Description of software implementation}
\label{implementation}

The OptiBounce algorithm requires an efficient method of solving the large scale nonlinear optimisation problem derived in the previous sections. In our prototype implementation this is made possible by IPOPT \cite{wachterImplementationInteriorpointFilter2006}. IPOPT uses a barrier method that iteratively solves a series of sub-problems indexed by a parameter $\mu$. As $\mu \rightarrow 0$, the series of partial solutions converges to the optimum of the original problem. For a generic optimisation problem:
\begin{equation}
\underset{\mathbf{x} \in \mathbb{R}^N}{\text{min}\ f(\mathbf{x})},\ \text{such that}\ c(\mathbf{x}) = 0,\ \mathbf{x} \geq 0,
\end{equation}
the barrier sub-problem is:
\begin{equation}
\underset{\mathbf{x} \in \mathbb{R}^N}{\text{min}\ \phi_{\mu}(\mathbf{x})},\ \phi_{\mu}(\mathbf{x}) \equiv f(\mathbf{x}) - \mu \sum_{i=1}^N \log(x^i),\ \text{such that}\ c(\mathbf{x}) = 0.
\end{equation}
The corresponding Karush–Kuhn–Tucker (KKT) optimality conditions are:
\begin{align}
\nabla f(\mathbf{x}) + \lambda \nabla c(\mathbf{x}) - \mathbf{z} &= 0, \\
c(\mathbf{x}) &= 0, \\
z^i &= \frac{\mu}{x^i}. 
\end{align}
At each step IPOPT uses an internal linear solver to find an approximate solution for these KKT conditions.	 The gradients $\nabla f$, $\nabla c$ must therefore be computed several times per iteration. In our case, this corresponds to taking the gradients of equations \ref{eq:continuity}, \ref{eq:collocation} and \ref{eq:v0_discrete} along with the objective function \ref{eq:objective_discrete} with respect to the state variables $\bm{\phi}_{k,j}$ and derivatives $\bm{u}_k$, $k = 0,...,N - 1$, $j = 0,...,d$. 

Since we use $N \approx O(100)$, computing gradients of the constraint and objective functions becomes an important computational cost. To do this as efficiently and accurately as possible, we make use of CasADi \cite{anderssonCasADiSoftwareFramework2019}. CasADi is a symbolic framework that provides differentiable, composable  primitives from which more complex functions can be constructed. Built-in automatic differentiation routines then allow for the calculation of gradients at a cost comparable to evaluation of the original function. This method is free of the discretisation error and instabilities associated with finite difference methods, and is much simpler to implement than symbolic differentiation. CasADi also comes with a native interface to IPOPT which uses highly optimised sparse matrix data types. This avoids redundant computations and ensures that large-but-sparse systems of the type used in OptiBounce can be solved efficiently, even in the case of hundreds or thousands of state variables. As demonstrated in section \ref{results}, this can be a highly efficient and accurate method for finding the bounce solution.

Another advantage of the CasADi-IPOPT stack is the ability to share computational load between different instances of related calculations. In this framework, solving an optimisation problem is split into two phases. In the first phase an in-memory symbolic representation of the problem is constructed, yielding an instance of \texttt{casadi::NLP}. In the second phase IPOPT is used to minimise the objective function. Importantly, CasADi allows the \texttt{casadi::NLP} instance to include unbound parameters. This means that for a given potential, the same object can be re-used to evaluate the bounce action at different temperatures. Since the setup cost can be comparable to or greater than the optimisation cost, this approach is particularly well suited to our design objective of quickly finding the nucleation temperature.

\section{Numerical Results}
\label{results}

\begin{figure}[htbp]
	\centering
	\includegraphics[width=0.5\textwidth]{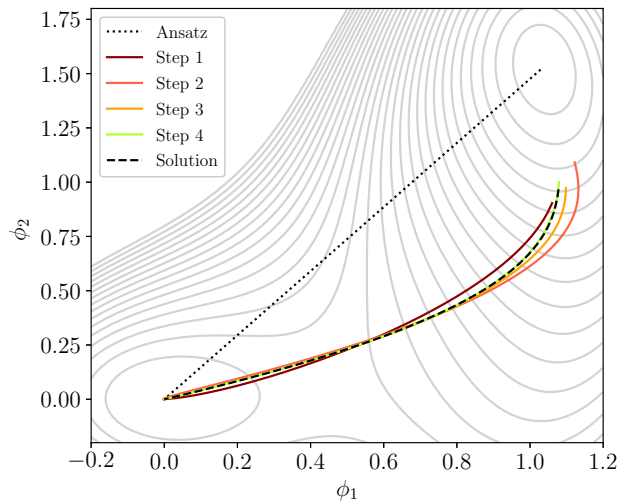}
	\caption{OptiBounce convergence on the two-field potential with $\delta = 0.3$. For this test case the \texttt{IPOPT} solver converges in nine steps. We show the initial straight-line ansatz, the first four steps, and the solution at step nine.}
	\label{fig:contours}
\end{figure}

\begin{figure}[htbp]
	\centering
	\begin{subfigure}[b]{0.5\textwidth}
		\centering
		 \includegraphics[width=\textwidth]{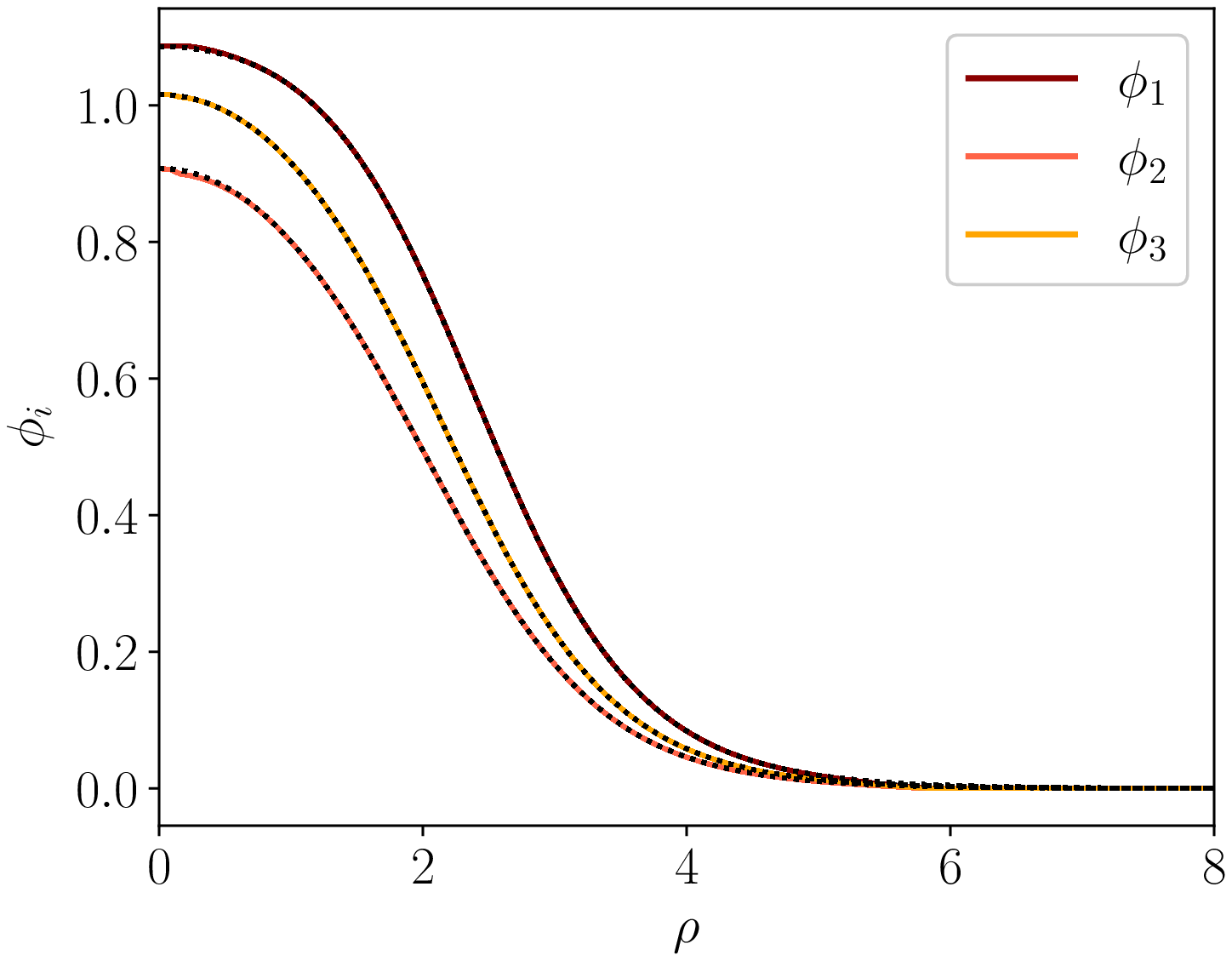}
        \caption{$\delta=0.3$}
	\end{subfigure}
	\begin{subfigure}[b]{0.5\textwidth}
		\centering
		 \includegraphics[width=\textwidth]{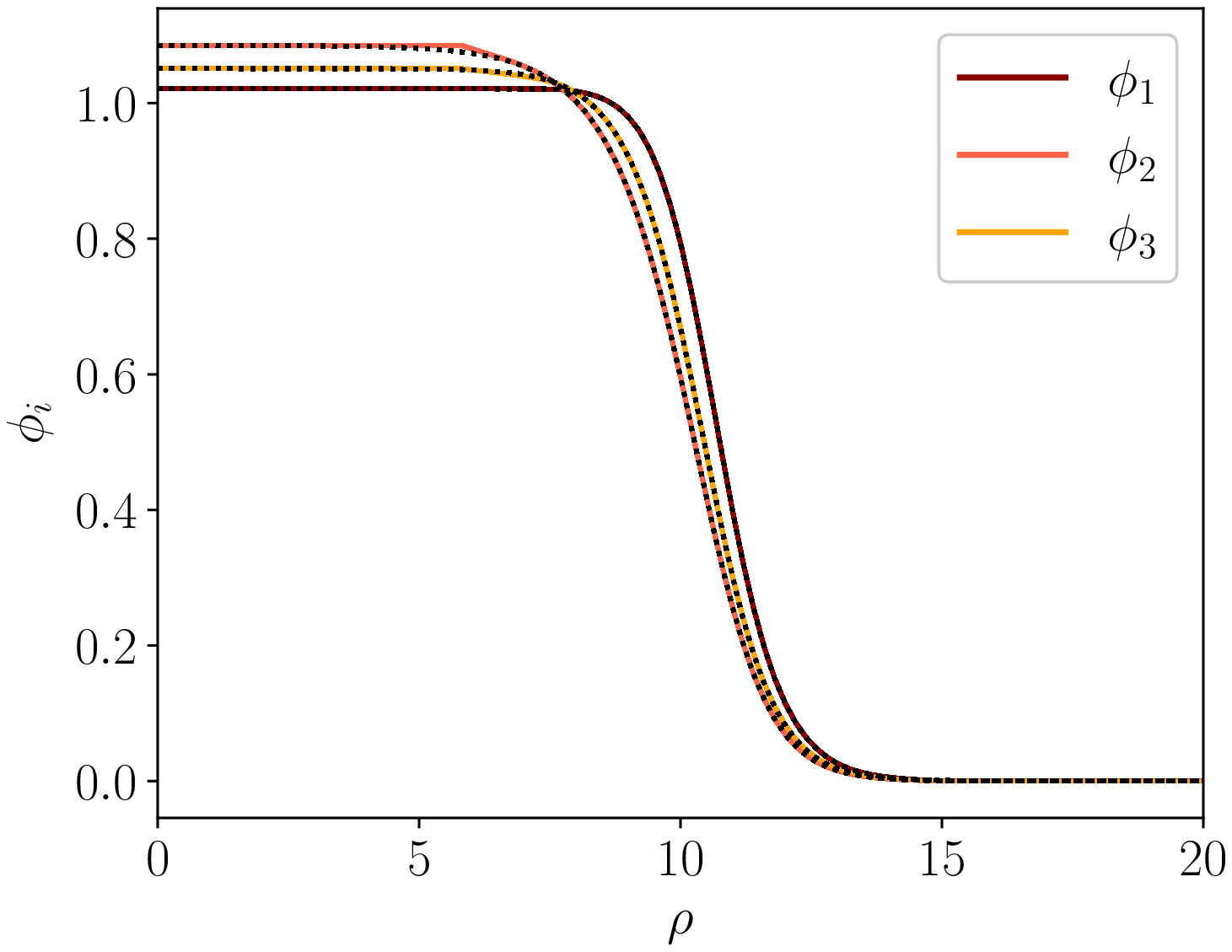}
        \caption{$\delta=0.03$}
	\end{subfigure}
	\caption{Thick- and thin-walled solutions for the test potential \ref{eq:test_potential} with $n_{\phi} = 3$. Solution curves are shown for the OptiBounce algorithm (dashed) and \texttt{FindBounce} (solid).}
	\label{fig:r_phi_thick_thin}
\end{figure}

Our objective in this section is to explore the performance and accuracy of the new algorithm on test potentials with multiple fields. We build our test cases using the form introduced in \cite{athronBubbleprofilerFindingField2019}:
\begin{equation}
\label{eq:test_potential}
V_{n_\phi} = \bigg(\bigg[\sum_{i=1}^{n_\phi} c_i (\phi_i - 1)^2 \bigg] -\delta\bigg)\bigg(\sum_{i=1}^{n_\phi} \phi_i^2 \bigg)
\end{equation}
where the coefficients $c_i$ are given in appendix \ref{coefficients}. The two field case with $\delta = 0.3$ is shown in figure \ref{fig:contours}. For $0 < \delta < 1$ this potential has a false vacuum at $(0,0)$ and a true vacuum in the vicinity of $(1,1)$. As $\delta \rightarrow 0$, the vacua approach degeneracy and the solution becomes thin-walled. In physical terms, this reflects the situation near the critical temperature when $(T - T_C)/T_C \ll 1$. Figure \ref{fig:r_phi_thick_thin} contrasts the different kinds of solutions obtained for large and small values of $\delta$.

%A common task when solving the bounce equation is to find the nucleation temperature. This is done by repeatedly solving the bounce equation while varying the temperature until the threshold condition \ref{eq:nucleation_approx} is satisfied. A strength of our algorithm is that these calculations can share the same \texttt{casadi::NLP} instance as described in section \ref{implementation}, paying the setup cost only once and varying the temperature between calls to IPOPT. 

\begin{figure}[htbp]
	\centering
	\includegraphics[width=0.5\textwidth]{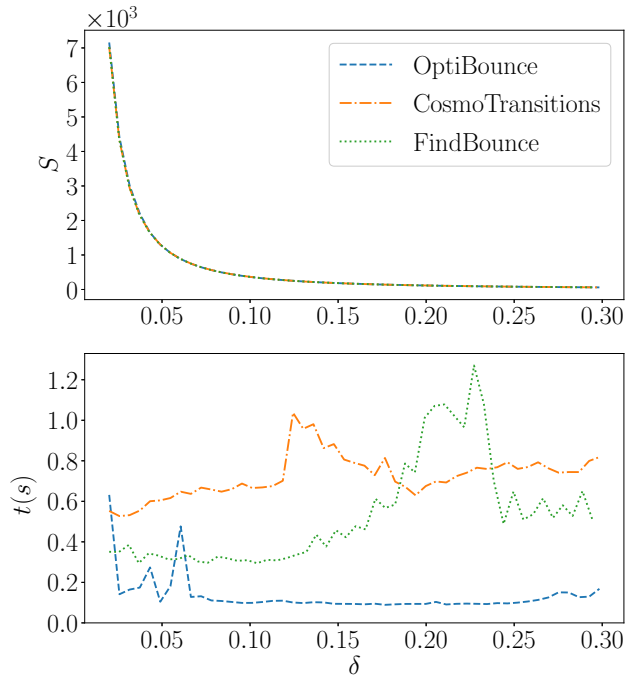}
	\caption{Action and timing data for the $n_{\phi} = 5$ test potential with $\delta \in [0.02,0.3]$. For OptiBounce the number of collocation points per element was increased from $d=3$ to $d=7$ to improve accuracy in the thin wall limit. Note that the OptiBounce timings do not include the one-off setup cost, as this computation was shared between all instances of the calculation.}
\label{fig:action_delta}
\end{figure}

As described in section \ref{implementation}, our algorithm is optimised for quickly finding the nucleation temperature. Figure \ref{fig:action_delta} models this kind of computation using the limit $\delta \rightarrow 0$ as a proxy for $T \rightarrow T_C$ on the $n_{\phi} = 5$ test potential. Away from the thin-wall limit, we find that the per-point computation time is greatly reduced compared to \texttt{CosmoTransitions} and \texttt{FindBounce}. For smaller values of $\delta$ the computation time becomes comparable to the other two codes. As noted in section \ref{discretisation}, this corresponds to the regime where the bubble wall scale is smaller than the grid size. We consider this to be a defect of the discretisation scheme rather than the algorithm as a whole, and expect that an adaptive scheme would allow better performance in the thin-wall limit. 

We also investigated the performance of the algorithm with larger numbers of fields up to $n_{\phi} = 20$. To ensure consistency between different values of $n_{\phi}$ we chose values of $\delta$ such that the action was in the range $S \in [200,400]$, representing cases away from both the thick- and thin-walled limits. Table \ref{table_benchmarks} presents the timings for each case along with comparisons to \texttt{CosmoTransitions} and \texttt{FindBounce}. In all cases the action obtained by the codes agrees to within $1\%$. For our algorithm the setup and solution times are reported separately. Considering the solution time only, we outperform the other codes for all values of $n_{\phi}$. If we also account for the setup time, for $n_{\phi} > 13$ \texttt{FindBounce} is faster on some points. However, we emphasise that when finding nucleation temperatures the setup time is a one-off cost which implies that for a sufficiently large number of evaluations our algorithm should be more efficient.

\begin{table*}[t]
\begin{ruledtabular}
\begin{tabular}{ ccccccccc }
$n_{\phi}$ & $\delta$ & Action (OB) & $\text{t}_{\text{setup}}$ (OB) & $\text{t}_{\text{sol}}$ (OB) & Action (FB) & $\text{t}_{\text{sol}}$ (FB) & Action (CT) & $\text{t}_{\text{sol}}$ (CT) \\
\hline
3 & 0.065 & 240.049 & 0.191 & 0.009 & 240.403 & 0.294 & 240.324 & 0.492 \\
4 & 0.11 & 227.023 & 0.217 & 0.081 & 230.864 & 0.615 & 230.397 & 3.223 \\
5 & 0.13 & 233.523 & 0.259 & 0.025 & 233.716 & 0.408 & 233.357 & 0.976 \\
6 & 0.15 & 270.363 & 0.300 & 0.031 & 270.578 & 0.442 & 271.769 & 3.917 \\
7 & 0.2 & 250.054 & 0.355 & 0.046 & 250.222 & 0.482 & 249.845 & 3.802 \\
8 &  0.22 & 268.259 & 0.404 & 0.085 & 268.486 & 0.606 & 269.368 & 4.143 \\
9 & 0.29 & 204.609 & 0.452 & 0.083 & 204.796 & 0.675 & 205.888 & 0.769 \\
10 & 0.27 & 261.468 & 0.507 & 0.088 & 261.703 & 0.779 & 261.273 & 1.279 \\
11 & 0.3 & 273.271 & 0.565 & 0.116 & 273.564 & 0.861 & - & - \\	
12 & 0.32 & 249.691 & 0.618 & 0.158 & 249.928 & 0.950 & - & - \\
13 & 0.39 & 293.383 & 0.671 & 0.153 & 293.653 & 1.012 & - & - \\
14 & 0.39 & 294.677 & 0.731 & 0.528 & 294.877 & 1.114 & - & - \\
15 & 0.42 & 312.760 & 0.816 & 0.455 & 313.039 & 1.222 & - & - \\
16 & 0.45 & 366.870 & 0.881 & 0.471 & 366.978 & 1.885 & - & - \\
17 & 0.52 & 342.537 & 0.961 & 0.734 & 342.893 & 1.395 & - & - \\
18 & 0.47 & 390.925 & 1.060 & 0.772 & 391.231 & 1.544 & - & - \\
19 & 0.56 & 339.287 & 1.116 & 1.537 & 339.467 & 2.436 & - & - \\
20 & 0.55 & 381.886 & 1.178 & 1.309 & 382.153 & 1.864 & - & - \\
\end{tabular}
\end{ruledtabular}
\caption{Timing data and action for test potentials with up to $n_{\phi} = 20$ fields. Results are shown for the OptiBounce algorithm, \texttt{CosmoTransitions} (CT) and \texttt{FindBounce} (FB). For OptiBounce we report the setup time $\text{t}_{\text{setup}}$ and solution time $\text{t}_{\text{sol}}$ separately, with the total runtime being $\text{t}_{\text{setup}} + \text{t}_{\text{sol}}$. \texttt{CosmoTransitions} results for $n_{\phi} > 10$ are not reported as the code issues an error when there are more than 10 fields.}	
\label{table_benchmarks}
\end{table*}

%Temp caption (notes): these are basically the same multifield potentials in the FB paper. However, I have tweaked the delta value for each to give an action in the range 200-400 (i.e, intermediate cases - not thin or thick walled) for the sake of consistency. Remember to mention that the interpolating polynomials are of degree 3, which is sufficient as the solutions vary smoothly.
\section{Conclusion}
\label{conclusion}

In this paper, we presented a new and efficient method for finding the bounce solution. The OptiBounce algorithm is based on Coleman, Glaser, and Martin's original ``reduced" action \cite{colemanActionMinimaSolutions1978}, which has a true minimum at a solution related to the bounce by a scale transform. We provided an analytic result showing that the scale transform is not even required to compute the bounce action (see equation \ref{eq:action}, appendix \ref{derivation}). To implement this method we designed a Legendre-Gauss collocation scheme to represent the bounce solution, yielding a large scale finite-dimensional optimisation problem over the collocation coefficients. Taking advantage of recent advances in large scale optimisation and automatic differentiation, we implemented the scheme using a CasADi-IPOPT stack to explore the accuracy and performance characteristics of OptiBounce. We were able to reproduce the results of the \texttt{CosmoTransitions} and \texttt{FindBounce} codes to within $O(1\%)$ on a set of test problems with 3-20 scalar fields. Our performance results suggest that the OptiBounce approach will be especially efficient for finding nucleation temperatures as the initial setup cost can be shared between executions. Neglecting this setup cost, our execution times are frequently orders of magnitude faster than the other codes we analysed. 

\begin{acknowledgements}
The author thanks Csaba Balazs and Peter Athron for their supervision, guidance, and support. Part of this work was completed at the École Polytechnique Fédérale de Lausanne during a research stay hosted by Professor Mikhail Shaposhnikov at the Laboratory for Particle Physics and Cosmology. The author thanks Professor Shaposhnikov for his generous hospitality. The work of M.B. was supported by an Australian Government Research Training Program (RTP) Scholarship and a Swiss Government Excellence Scholarship from the Federal Commission for Scholarships for Foreign Students (FCS), with supplementary funding from ERC-AdG-2015 Grant No. 694896.
\end{acknowledgements}

\appendix	
\section{Derivation of optimisation algorithm}
\label{derivation}
In this section we provide a derivation establishing that the algorithm described in section \ref{algorithm} recovers the bounce action and field profile via equations \ref{eq:action} and \ref{eq:lambda}. Our starting point is the ``reduced problem" defined by Coleman, Glaser, and Martin \cite{colemanActionMinimaSolutions1978}. Recall that the set $\mathbf{\Phi}$ contains all field profiles satisfying the boundary conditions $\dot{\phi}_i(0) = 0$, $\lim_{\rho\to\infty}\phi_i(\rho) = 0$, $1 \leq i \leq n_{\phi}$. For each $\bm{\phi} \in \mathbf{\Phi}$, we consider the scale transformation $\bm{\phi}_{\sigma}(\rho) = \bm{\phi}(\sigma^{-1}\rho)$ for some $\sigma > 0$. Firstly, the action transforms as:
\begin{equation}
\label{eq:scaling}
S_E[\bm{\phi}_{\sigma}(\rho)] = \sigma^{D-2}T[\bm{\phi}(\rho)] + \sigma^D V[\bm{\phi}(\rho)].
\end{equation}
Since the bounce solution $\bm{\phi}^B$ makes $S_E$ stationary, this variation should vanish around $\sigma = 1$:
\begin{equation}
\frac{dS_E	[\bm{\phi}_{\sigma}^B(\rho)]}{d\sigma}\bigg|_{\sigma=1} = (D - 2)T[\bm{\phi}^B(\rho)] + D V[\bm{\phi}^B(\rho)] = 0,
\end{equation}
yielding the relation
\begin{equation}
\label{eq:t_and_v}
V = \frac{2 - D}{D}T.
\end{equation}
Since $T > 0$, for $D \geq 2$ this implies $V[\bm{\phi}^B] < 0$. Moreover, if the bounce solution exists then the level set:
\begin{equation}
\bm{\Phi}_{V_0} \equiv \{\bm{\phi} \in \bm{\Phi} : V[\bm{\phi}] = V_0\}
\end{equation}
is not empty for any $V_0 < 0$ since clearly $V[\bm{\phi}^B_{\sigma}] = V_0$ for some $\sigma > 0$. Therefore the minimizer:  
\begin{equation}
\bm{\phi}^{*} \equiv \argmin_{\bm{\phi} \in \Phi_{V_0}} T[\bm{\phi}(\rho)]
\end{equation}
must exist. If we implement the constraint $V[\bm{\phi}] = V_0$ with a Lagrange multiplier $\lambda$, $\bm{\phi}^*$ is a stationary point of the augmented Lagrangian:   
\begin{equation}
S_E[\bm{\phi}(\rho),\lambda] = T[\bm{\phi}(\rho)] + \lambda(V[\bm{\phi}(\rho)] - V_0).
\end{equation}
In fact, if we relax the constraint and introduce the optimal Lagrange multiplier $\lambda_*$, it is also a stationary point of:
\begin{align}
S_{\lambda_*}[\bm{\phi}(\rho)] &= T[\bm{\phi}(\rho)] + \lambda_* V[\bm{\phi}(\rho)] \\
&= \int_0^{\infty} d\rho \rho^{D-1}\bigg[\frac{1}{2}|\dot{\bm{\phi}}|^2 + \lambda_* V(\bm{\phi})\bigg],
\end{align}
and so has equations of motion:
\begin{equation}
\ddot{\phi^*_i} + \frac{D - 1}{\rho}\dot{\phi^*_i} = \lambda_* \frac{\partial V}{\partial \phi^*_i}. 
\end{equation}
This means that we can recover the bounce solution by a scale transform  $\bm{\phi}^B(\rho) = \bm{\phi}^* (\rho/\sqrt{\lambda_*})$, since then:
\begin{equation}
\ddot{\phi^B_i} + \frac{D - 1}{\rho}\dot{\phi^B_i} = \frac{\partial V}{\partial \phi^B_i}. 
\end{equation}
Moreover, since $\bm{\phi}^B$ is a stationary point of $S_E[\bm{\phi}(\rho)]$ we can directly obtain the action by inverting equation \ref{eq:t_and_v}:
\begin{align}
S_E[\bm{\phi}^B(\rho)] &= \frac{2}{2-D} V[\bm{\phi}^*(\rho/\sqrt{\lambda_*})] \\
&= \frac{2}{2 - D}\lambda_*^{\frac{D}{2}} V[\bm{\phi}^*(\rho)] \\
&= \frac{2 \lambda_*^{\frac{d}{2}} V_0}{2 - D}.
\end{align}
Alternatively, from equation \ref{eq:scaling} we have:
\begin{equation}
T[\bm{\phi}^B(\rho)] = \lambda_{*}^{1 - \frac{D}{2}} T[\bm{\phi}^B(\rho/\sqrt{\lambda_*})] \equiv \lambda_{*}^{1 - \frac{D}{2}} T_0.
\end{equation}
Inserting equation \ref{eq:t_and_v} into $S = T + V$ then gives:
\begin{equation}
S_E[\bm{\phi}^B(\rho)] = \frac{2}{D}T[\bm{\phi}^B(\rho)] = \frac{2 \lambda_{*}^{\frac{D}{2} - 1} T_0}{D}.
\end{equation}
Equality between the two expressions for $S_E[\bm{\phi}^B(\rho)]$ means we can write $\lambda_*$ in terms of $T_0/V_0$:
\begin{equation}
\label{eq:lambda_star}
\lambda_* = \bigg(\frac{2 - D}{D}\bigg) \frac{T_0}{V_0}.
\end{equation}
This means that we can express $S_E[\bm{\phi}^B(\rho)]$ in terms of $T_0$, $V_0$ and $D$ only:
\begin{equation}
S[\bm{\phi}(\rho)] = \bigg[\bigg(\frac{2 - D}{D}\bigg) \frac{T_0}{V_0}\bigg]^{\frac{D}{2} - 1} \frac{2 T_0}{D}.
\end{equation}

%\section{Coefficients for multi field potentials}
\label{coefficients}
\begin{table*}
\centering
\begin{ruledtabular}
\begin{tabular}{ cll }
$n_{\phi}$ & $\delta$ & $c_i$ \\
\hline
3 & 0.065 & 0.684373, 0.181928, 0.295089 \\
4 & 0.11 & 0.534808, 0.77023, 0.838912, 0.00517238 \\
5 & 0.13 & 0.4747, 0.234808, 0.57023, 0.138912, 0.517238 \\
6 & 0.1 & 0.34234, 0.4747, 0.234808, 0.57023, 0.138912, 0.517238 \\
7 & 0.2 & 0.5233, 0.34234, 0.4747, 0.234808, 0.57023, 0.138912, 0.517238 \\
8 & 0.22 & 0.2434, 0.5233, 0.34234, 0.4747, 0.234808, 0.57023, 0.138912, 0.51723 \\
9 & 0.29 & 0.21, 0.24, 0.52, 0.34, 0.47, 0.23, 0.57, 0.14, 0.52 \\
10 & 0.27 & 0.12, 0.21, 0.24, 0.52, 0.34, 0.47, 0.23, 0.57, 0.14, 0.52 \\
11 & 0.3 & 0.23, 0.21, 0.21, 0.24, 0.52, 0.34, 0.47, 0.23, 0.57, 0.14, 0.52 \\
12 & 0.32 & 0.12, 0.11, 0.12, 0.21, 0.24, 0.52, 0.34, 0.47, 0.23, 0.57, 0.14, 0.52 \\
13 & 0.39 & 0.54, 0.47, 0.53, 0.28, 0.35, 0.27, 0.42, 0.59, 0.33, 0.16, 0.38, 0.35, 0.17 \\
14 & 0.39 & 0.39, 0.23, 0.26, 0.40, 0.11, 0.42, 0.41, 0.27, 0.42, 0.54, 0.18, 0.59, 0.13, 0.29 \\
15 & 0.42 & 0.21, 0.22, 0.22, 0.23, 0.39, 0.55, 0.43, 0.12, 0.16, 0.58, 0.25, 0.50, 0.45, 0.35, 0.45 \\
16 & 0.45 & 0.42, 0.34, 0.43, 0.22, 0.59, 0.41, 0.58, 0.41, 0.26, 0.45, 0.16, 0.31, 0.39, 0.57, 0.43, 0.10 \\
17 & 0.52 & 0.24, 0.35, 0.39, 0.56, 0.37, 0.41, 0.52, 0.31, 0.52, 0.22, 0.58, 0.39, 0.39, 0.17, 0.46, 0.30, 0.37 \\
18 & 0.47 & 0.18, 0.17, 0.30, 0.22, 0.38, 0.48, 0.11, 0.49, 0.43, 0.47, 0.21, 0.29, 0.32, 0.36, 0.30, 0.56, 0.46, 0.42 \\
19 & 0.56 & 0.40, 0.14, 0.10, 0.43, 0.39, 0.27, 0.33, 0.59, 0.48, 0.36, 0.24, 0.28, 0.51, 0.59, 0.40, 0.39, 0.24, 0.35, 0.20 \\
20 & 0.55 & 0.42, 0.11, 0.47, 0.13, 0.16, 0.24, 0.58, 0.53, 0.38, 0.44, 0.18, 0.46, 0.47, 0.27, 0.53, 0.24, 0.33, 0.40, 0.32, 0.29 \\
\end{tabular}
\end{ruledtabular}
\caption{Coefficients used to generate the timing results in table \ref{table_benchmarks} with the potential defined in equation \ref{eq:test_potential}. For $3 \leq n_{\phi} \leq 8$, the $c_i$ are taken from \cite{athronBubbleprofilerFindingField2019}, with the remainder for $n_{\phi} \geq 9$ from \cite{guadaFindBouncePackageMultifield2020a}.} 
\end{table*}

%\printbibliography
\clearpage

\bibliography{algorithm_paper}

\end{document}